\pdfoutput=1
\documentclass[proceedings, preprint]{rmaa}

\usepackage{paralist}

\usepackage{psfrag,color}

\SetYear{2014}
\SetConfTitle{ADeLA}

\title{Bright Star Astrometry with URAT} 

\author{
  N. Zacharias\altaffilmark{1} }

\altaffiltext{1}{U.S.~Naval Observatory, 3450 Mass.~Ave.~NW,
  Washington DC, 20392, USA (nz@usno.navy.mil).}

\shortauthor{Zacharias}
\shorttitle{URAT Bright Star Astrometry}

\listofauthors{N. Zacharias}
\indexauthor{Zacharias, N.}

\abstract{The U.S.~Naval Observatory Robotic Astrometric Telescope 
  (URAT) is observing the northern sky since April 2012 for an
  astrometric survey.  Multiple overlaps per year are performed
  in a single bandpass (680$-$750 nm) using the ``redlens" 20 cm aperture
  astrograph and a mosaic of large CCDs.  Besides the regular, deep
  survey to magnitude 18.5, short exposures with an objective grating
  are taken to access stars as bright as 3rd magnitude.
  A brief overview of the program, observing and reductions is given.  
  Positions on the 8 to 20 mas level are obtained of 
  66,202 Hipparcos stars at current epochs.  These are compared
  to the Hipparcos Catalog to investigate its accuracy.
  About 20\% of the observed Hipparcos stars are found to have
  inconsitent positions with the Hipparcos Catalog prediction
  on the 3 sigma level or over (about 75 mas or more discrepant
  position offsets).  Some stars are now seen at an arcsec
  (or 25 sigma) off their Hipparcos Catalog predicted position.}

\addkeyword{Astrometry}
\addkeyword{Survey}
\addkeyword{Stars: Hipparcos Catalog}

\begin{document}
\maketitle

\section{Introduction}

Since April 2012 the U.S.~Naval Observatory (USNO) is conducting
the USNO Robotic Astrometric Telescope (URAT) survey.
The goal of this project is to establish a deep (18+ mag),
very accurate (10 mas level), optical reference frame based
on the Hipparcos / ICRF system using UCAC4 reference stars.
The program will also identify nearby stars unbiased by proper 
motion selection and obtains accurate positions of bright stars
at current epochs, which is the topic of this paper.
Using the global coordinate system of Hipparcos through the UCAC4 
reference stars, the URAT observations nevertheless give accurate 
positions of individual Hipparcos stars largely independent of the
Hipparcos catalog.
The URAT observations of these bright stars are compared to the
Hipparcos Catalog to assess the accuracy of the Hippacos
Catalog positions and proper motions and to identify discrepancies,
i.e.~stars which in reality are not at the position predicted by
the Hipparcos Catalog.

\begin{figure}[!t]
  \includegraphics[angle=0,width=\columnwidth]{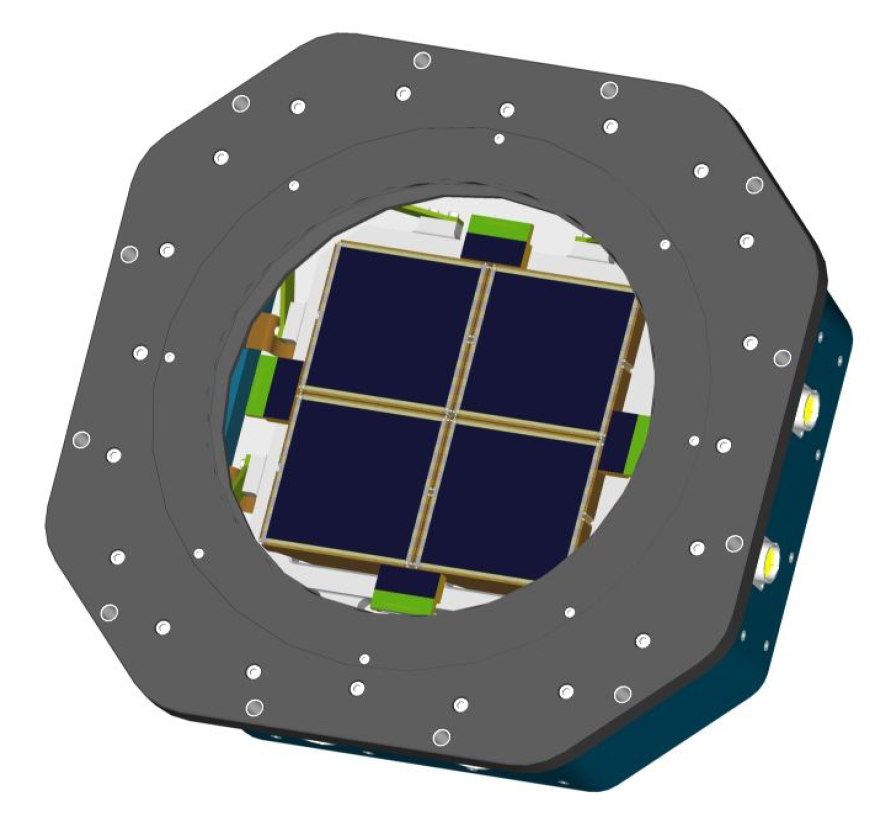}
  \caption{URAT focal plane assembly design by Semiconductor
     Technology Associates (STA).  
     The window has a clear aperture of 300 mm.  
     The 4 large CCDs have 111 million pixels each.}
  \label{fig:simple}
\end{figure}

\section{Instrument}

The ``redlens" of the USNO astrograph is now fully utilized
with its new, large focal plane of 286 mm diameter using a mosaic
of 4 STA1600 CCDs (Fig.~1).  Each CCD has 10,560 by 10,560 pixels and 
covers about 7 square deg of sky at 0.9 arcsec/pixel resolution.
The dewar window serves as filter for the fixed 680$-$750 nm bandpass.
A completely new tube structure was designed and
built by the USNO instrument shop in Washington DC.
Observations are performed at NOFS.
For more details see Zacharias et al.~(2012, 2015) and the URAT
homepage\footnote{www.usno.navy.mil/usno/astrometry/optical-IR-prod/urat}.

\section{Observations}

\begin{figure}[!t]
  \includegraphics[angle=0,width=\columnwidth]{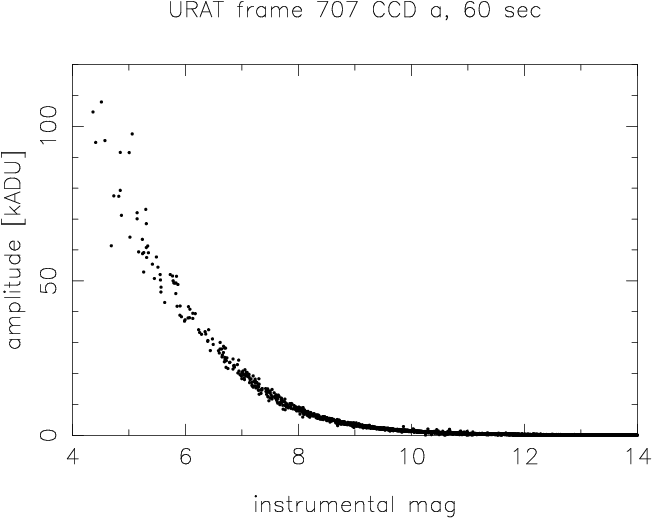}
  \caption{Fit amplitude of individual star images as a function of
     instrumental magnitude for an example of URAT exposure and CCD.}
  \label{fig:simple}
\end{figure}

\begin{figure}[!t]
  \includegraphics[angle=0,width=\columnwidth]{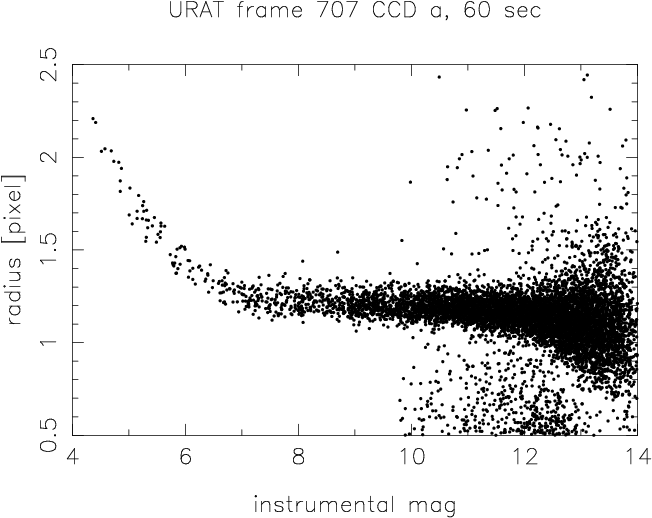}
  \caption{Fit radius of individual star images as a function of
     instrumental magnitude for an example of URAT exposure and CCD.}
  \label{fig:simple}
\end{figure}

\begin{figure}[!t]
  \includegraphics[angle=0,width=\columnwidth]{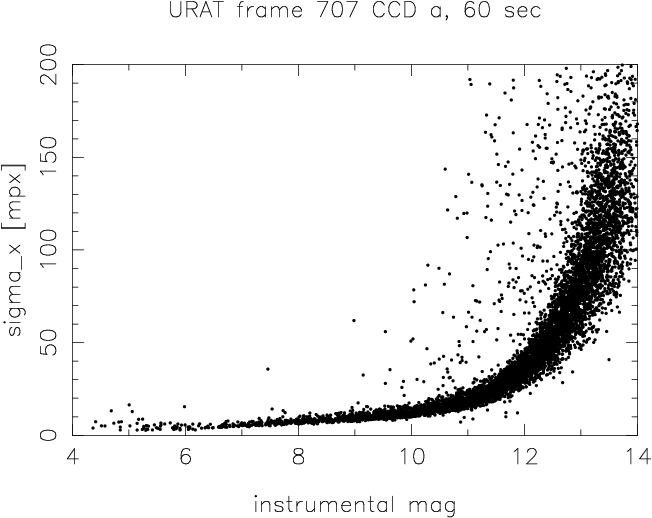}
  \caption{Position fit error of individual star images as a function of
     instrumental magnitude for an example of URAT exposure and CCD.}
  \label{fig:simple}
\end{figure}

For the regular survey each field is observed with a 240 and a 60 sec
exposure.  During the week of full Moon, short exposures (10 or 20) and
30 sec are taken with an objective grating which provides diffraction
images about 5 magnitudes fainter than the central image.
Another about 2 magnitudes dynamical range is gained by the clocked
anti-blooming (CAB) feature of the CCDs.  
Fig.~2 shows the image profile fit amplitude versus instrumental
magnitude.  Traditional saturation is reached around 30k ADU,
thus at about instrumental magnitude 6.7.
Beyond saturation the image profiles 
get wider (Fig.~3), however, the image fit position error (Fig.~4)
remains low (10 milli-pixel or less), up to about 2.5 magnitudes
brighter than saturation.
Beyond that systematic errors become too large for sufficient calibration.
Typical 10 and 30 sec exposures saturate around calibrated URAT magnitude 
9 and 10, respectively.  The 1st order grating images of these exposures
saturate around magnitude 4 and 5, respectively.

Multiple exposures per area of sky were taken with diagonal shifts
of field center to allow the same star fall onto different parts of
the CCD and different CCDs.  Typically 10 exposures per star and
observing run of several nights duration were obtained.

\section{Reductions}

\begin{figure}[!t]
  \includegraphics[bb= 39 123 534 640,clip,width=1.0\columnwidth]
     {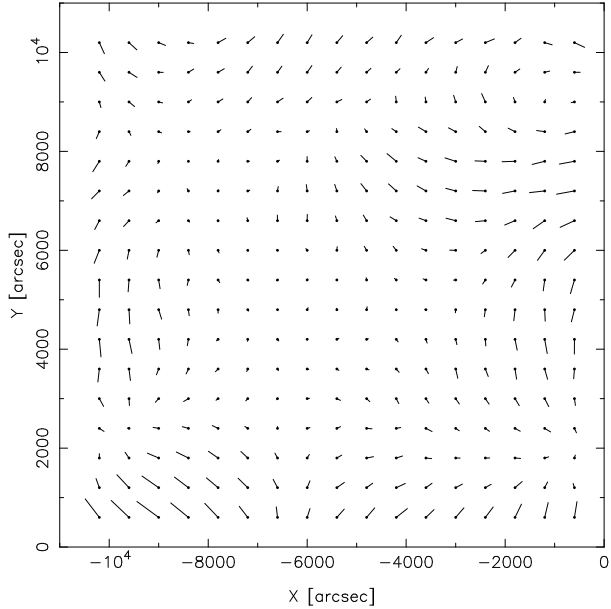}
  \caption{Example of a field distortion pattern, derived from
     URAT exposures of a single night for CCD B. The vectors are
     scaled by a factor of 5000.} 
  \label{fig:simple}
\end{figure}

Astrometric reductions of all data were performed using UCAC4 reference
stars.  A single CCD exposure typically has between 500 and 5000 such
reference stars, thus ``averaging out" systematic zonal errors
of the UCAC4 or Tycho-2 over 3 degrees.
An 8-parameter ``plate" model (linear + tilt terms) was adopted.
The 3rd order optical distortion of the lens is too small to require such 
a term in the model.  However, a general field distortion pattern was 
constructed and applied for each CCD separately, which takes out 
systematic errors as a function of $x,y$ pixel location due to residual 
distortions, e.g.,~caused by the filter and lens (Fig.~5).

\begin{figure}[!t]
  \includegraphics[angle=0,width=1.2\columnwidth]{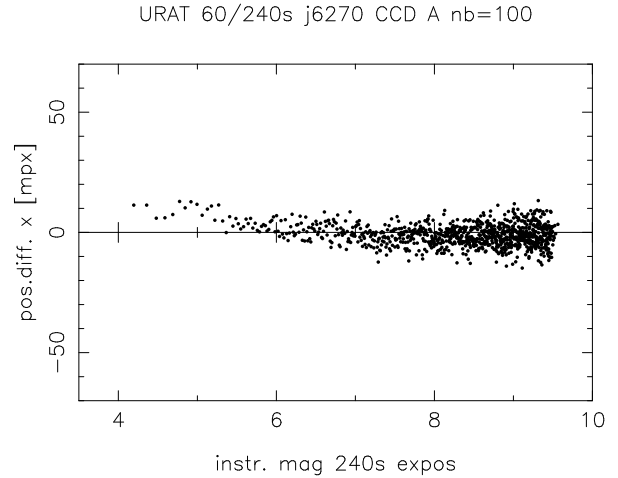}
  \caption{Position differences along $x$ between 240 and 60 sec exposures of
     the same field, averaged over all exposure pairs of night j6270
     for CCD A.}
  \label{fig:simple}
\end{figure}

\begin{figure}[!t]
  \includegraphics[angle=0,width=1.2\columnwidth]{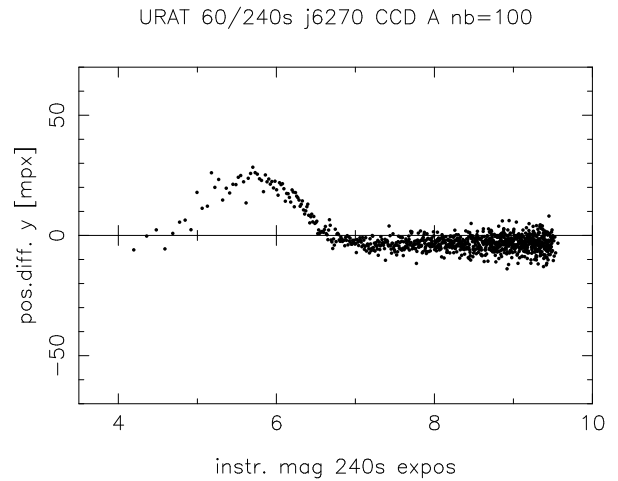}
  \caption{Same as previous figure for $y$ coordinate.}
  \label{fig:simple}
\end{figure}

The $x,y$ data were also corrected for the pixel phase error,
a sine-wave as a function of sub-pixel coordinate and amplitude of
about 5 to 15 mas, depending on the width of image profiles.
Systematic position errors (separately for $x$ and $y$ coordinates and CCD) 
of saturated images were found to depend on time.  The data were split
into groups by epoch and separate corrections derived and applied.
Examples are shown in Figs.~6 and 7.
These corrections are typically in the range of 10 to 100 mas.
Individual positions of stars (per exposure, CCD and grating order)
were combined to weighted mean positions for all observations of
an individual run (observing period around a given full Moon).
These positions were then compared to the Hipparcos Catalog 
(van Leeuwen 2007)
positions predicted for the URAT mean observing epoch utilizing
the Hipparcos Catalog's proper motions and parallaxes.

\begin{figure}[!t]
  \includegraphics[angle=0,width=1.2\columnwidth]{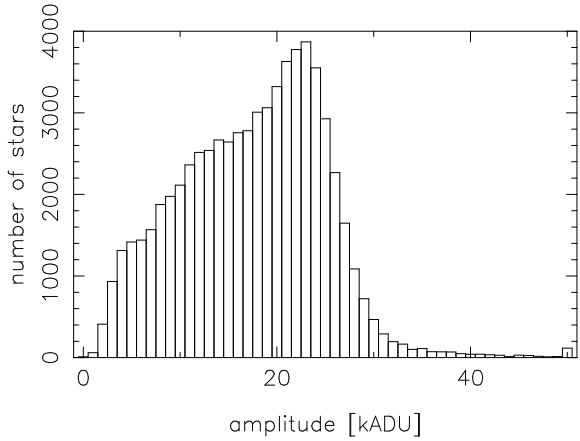}
  \caption{Distribution of amplitudes of URAT images.}
  \label{fig:simple}
\end{figure}

\begin{figure}[!t]
  \includegraphics[angle=0,width=1.2\columnwidth]{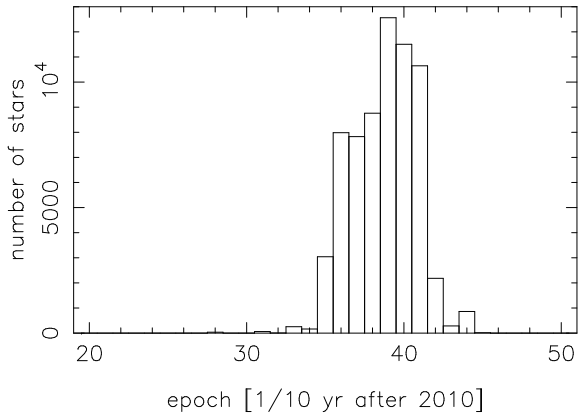}
  \caption{Distribution of URAT observational epochs.}
  \label{fig:simple}
\end{figure}

\begin{figure}[!t]
  \includegraphics[angle=0,width=1.2\columnwidth]{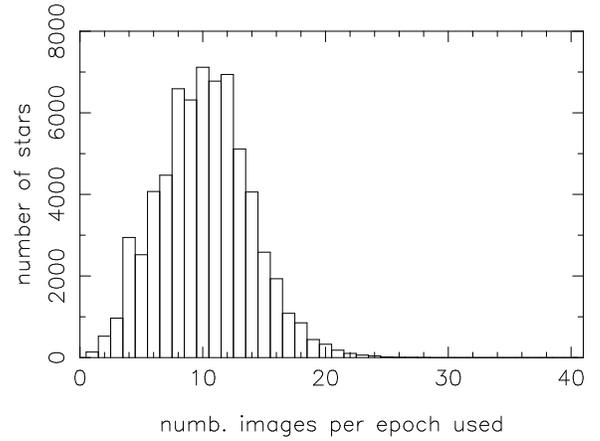}
  \caption{Distribution of number of observations per observing run
   of several nights around a full Moon.}
  \label{fig:simple}
\end{figure}

\section{Results}

\begin{figure}[!t]
  \includegraphics[angle=0,width=1.2\columnwidth]{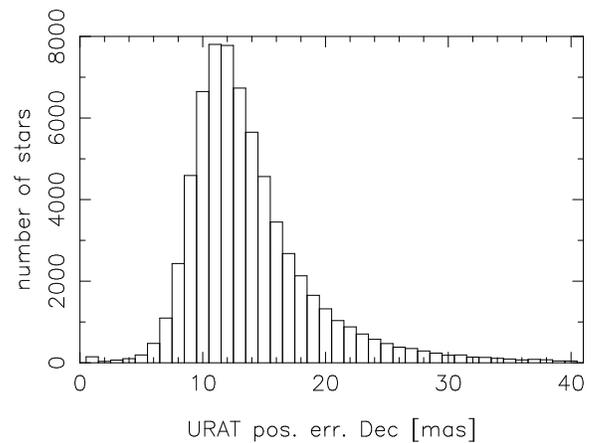}
  \caption{Distribution of URAT position errors (Dec).}
  \label{fig:simple}
\end{figure}

\begin{figure}[!t]
  \includegraphics[angle=0,width=1.2\columnwidth]{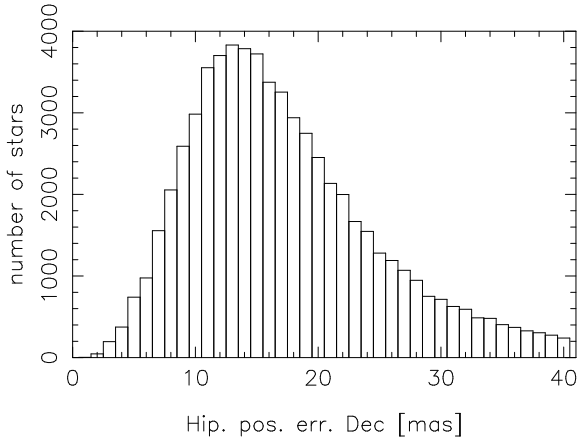}
  \caption{Distribution of Hipparcos position errors at URAT epoch (Dec).}
  \label{fig:simple}
\end{figure}

\begin{figure}[!t]
  \includegraphics[angle=0,width=1.2\columnwidth]{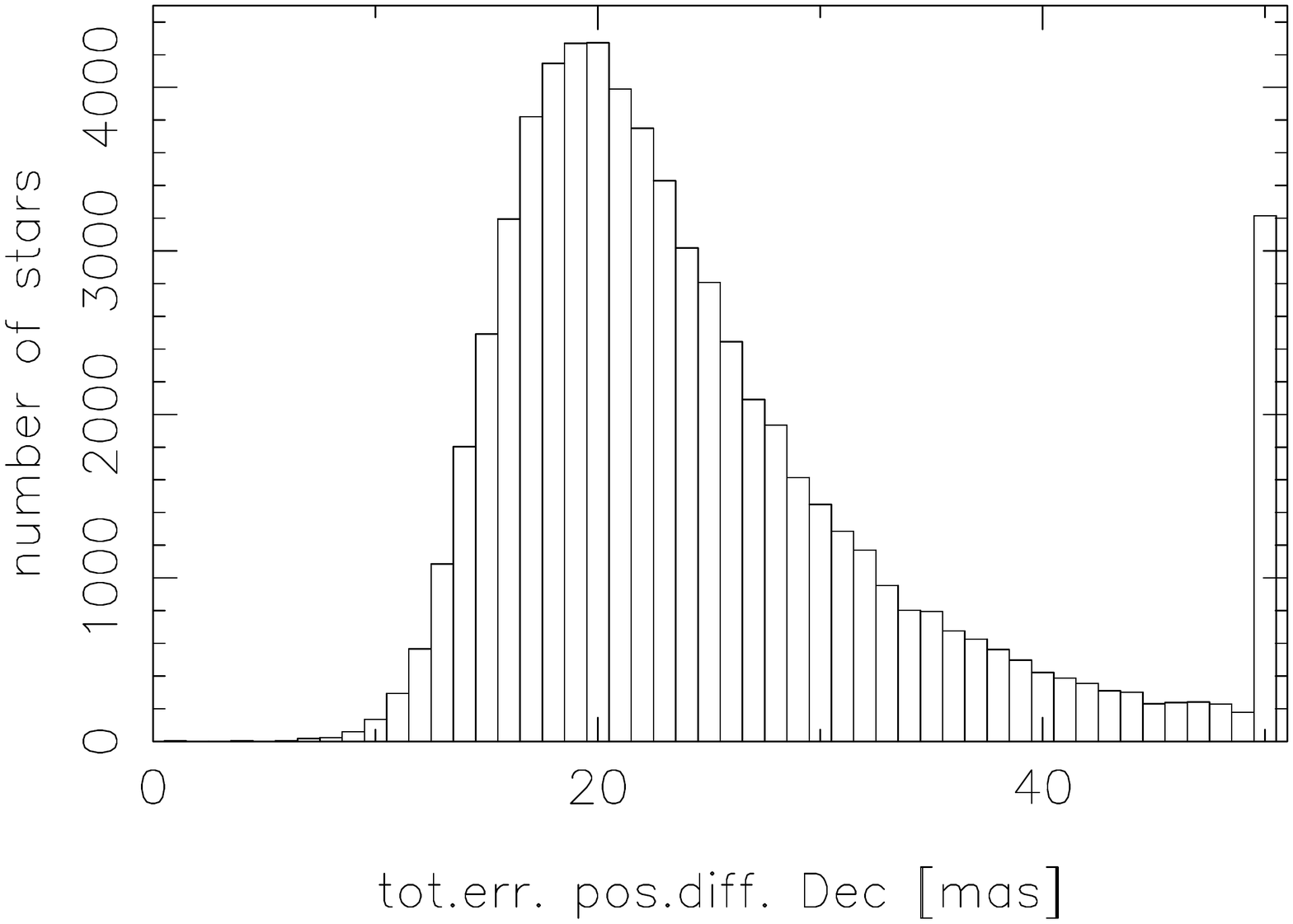}
  \caption{Distribution of URAT-Hipparcos position difference errors (Dec).}
  \label{fig:simple}
\end{figure}

Over 29,000 exposures from 85 nights of 17 runs (epoch groups) with 
grating observations between April 2012 and June 2014 were used for 
this investigation.  These URAT observations cover almost all sky 
between declinations $-5^{\circ}$ and $+89^{\circ}$ and contain over
a billion individual positions.
Among these, 66,202 Hipparcos stars were identified, the subject of
this investigation.

The distribution of the URAT observed mean image profile amplitudes
of the Hipparcos stars is shown in Fig.~8.
Thus most data are not saturated (under 30k ADU) and only for a
small fraction of stars the CAB regime is used.
Fig.~9 shows the distribution of epochs of these URAT observations,
which thus have an epoch difference of about 22.5 years to the original
Hipparcos mean observing epoch (1991.25).
The mean number of URAT observations per star and observing
run of several nights is shown in Fig.~10.

\begin{figure}[!t]
  \includegraphics[angle=0,width=1.2\columnwidth]{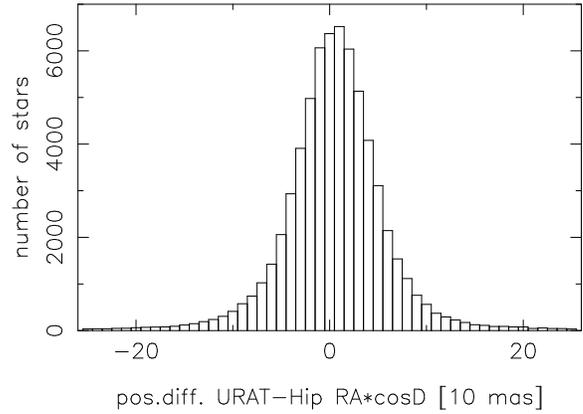}
  \caption{Distribution of URAT-Hipparcos position differences (RA*cosDec).}
  \label{fig:simple}
\end{figure}

\begin{figure}[!t]
  \includegraphics[angle=0,width=1.2\columnwidth]{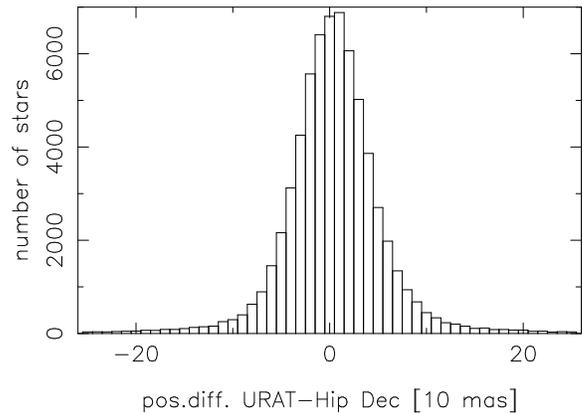}
  \caption{Distribution of URAT-Hipparcos position differences (Dec).}
  \label{fig:simple}
\end{figure}

Figs.~11 and 12 show the distribution of URAT and Hipparcos positional
errors at the URAT epoch, respectively.  Results for the declination
component are shown, which are similar to the results along RA.
Both data sets are of comparable precision at current epochs.
Fig.~13 shows the RMS combined error which is to be used as error 
for the position differences to be looked at next.

The distributions of the URAT$-$Hipparcos position differences are
shown in Figs.~14 and 15 for the RA and Dec components, respectively.
Fig.~16 shows the same as Fig.~15, however, normalized by the total,
combined error of the position difference (i.e.~ in ``sigmas").
It is obvious that a large fraction of stars show much larger
position differences than predicted from Gaussian statistics.
Fig.~17 shows a zoom-in of Fig.~16 to highlight the ``outliers".

\begin{figure}[!t]
  \includegraphics[angle=0,width=1.2\columnwidth]{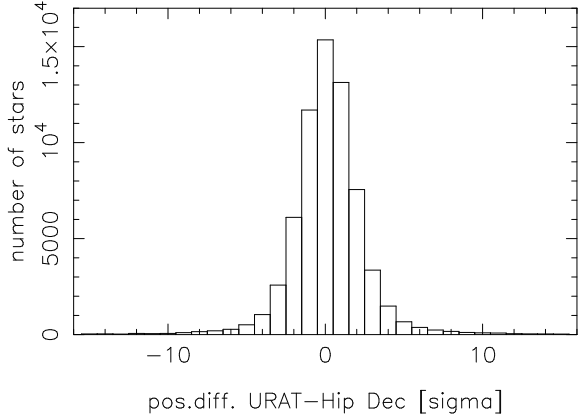}
  \caption{Distribution of URAT-Hipparcos normalized position 
     differences (Dec).}
  \label{fig:simple}
\end{figure}

\begin{figure}[!t]
  \includegraphics[angle=0,width=1.2\columnwidth]{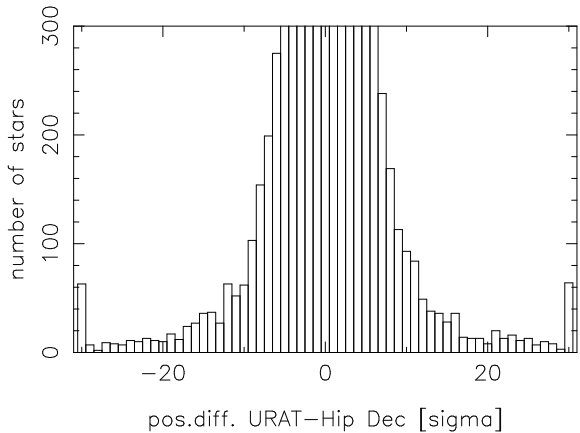}
  \caption{Zoom of previous figure to highlight large position differences.}
  \label{fig:simple}
\end{figure}

A quantitative evaluation of the large number of large URAT$-$Hipparcos
position differences is provided in Tables.~1 and 2.
The discrepancy between the current epoch actually observed positions 
of those stars and their predicted positions based on the Hipparcos 
Catalog (re-reduction version of 2007) is likely due to the correlations 
of parallax, proper motion and possible orbital motions estimated from a
very short time span (3.5 years) of Hipparcos observations.

\begin{table}[h]
\caption{Statistics of large URAT$-$Hipparcos position differences [mas].}
\begin{center}
\begin{tabular}{rrr}
\hline
 pos.diff.  & number  & percentage \\
 larger than& of      &   of   \\
  (mas)     & stars   &  stars \\
\hline
   1000     &   255 &  0.4 \\
    500     &   778 &  1.2 \\
    400     &  1087 &  1.6 \\
    300     &  1573 &  2.4 \\
    200     &  2651 &  4.0 \\
    150     &  3943 &  6.0 \\
    100     &  7569 & 11.4 \\
     75     & 13085 & 19.8 \\
\hline
\end{tabular}
\end{center}
\end{table}

\begin{table}[h]
\caption{Statistics of large URAT$-$Hipparcos position differences [sigma].}
\begin{center}
\begin{tabular}{rrr}
\hline
 pos.diff.  & number  & percentage \\
 larger than& of      &   of   \\
 ... sigma  & stars   &   stars   \\
\hline
    25.0  &   298  &  0.5 \\
    15.0  &   759  &  1.1 \\
    10.0  &  1489  &  2.2 \\
     8.0  &  2106  &  3.2 \\
     6.0  &  3387  &  5.1 \\
     5.0  &  4633  &  7.0 \\
     4.0  &  7240  & 10.9 \\
     3.0  & 13458  & 20.3 \\
\hline
\end{tabular}
\end{center}
\end{table}

\section{Conclusions}

The large amount of highly accurate observations of bright stars
by the URAT program allows a check on the accuracy of Hipparcos
Catalog positions of individual stars at current epochs. 
With typical precisions of 8 to 20 mas the URAT observations are
at least as precise as the Hipparcos positions at mean epoch of 2014.
Some discrepancies on the order of 25 sigma or arcsecond level are seen.
The fraction of stars with position differences exceeding 3-sigma 
(75 mas) of combined URAT and Hipparcos formal positional errors
is about 20\% for our data sample of over 66,000 Hipparcos stars,
mainly on the northern hemisphere.

\vspace*{2mm}
\noindent
{\bf Acknowledgement:} The author thanks the entire URAT team.
Please see the URAT1 release paper and readme file.

\center{REFERENCES} 
\begin{flushleft}
van Leeuwen, F. 2007, Springer Science Library, Vol.~350
\vspace*{2mm}

Zacharias, N., Bredthauer, G., DiVittorio, M., Finch, C., Gaume, R.,
Harris, F., Rafferty, T., Rhodes, A., Schultheis, M., Subasavage, J,
Tilleman, T., Wieder, G.  2012
   "The URAT project"
   presentation at IAU GA 28, Com.8 science session
   http://www.ast.cam.ac.uk/ioa/iau\_comm8/iau28/
\vspace*{2mm}

Zacharias, N. et al. 2015, URAT1 catalog release, submitted to AJ.
\end{flushleft}

\end{document}